%% file: main.tex
\documentclass[conference]{IEEEtran}
\IEEEoverridecommandlockouts
\usepackage{authblk}
\usepackage{cite}
\usepackage{amsmath,amssymb,amsfonts}
\usepackage{adjustbox}
\usepackage{algorithmic}
\usepackage{caption}
\usepackage{multicol}
\usepackage{import}
\usepackage{algorithmic}
\usepackage{graphicx}
\usepackage[colorlinks,
    linkcolor=blue,
    citecolor=blue,
    pdftex,
    pdfauthor={assert},
    pdftitle={autographql},
    pdfsubject={unicorns and apis},
    pdfkeywords={graphql, execution},
    pdfproducer={mel brooks},
    pdfcreator={gru}]{hyperref}
\usepackage[dvipsnames, table]{xcolor}
\usepackage{tabularx}
\usepackage{textcomp}
\usepackage{xcolor}
\usepackage{xspace}
\usepackage{booktabs}
\definecolor{light-gray}{gray}{0.95}
\definecolor{pgreen}{RGB}{5,205,107}
\definecolor{pblue}{RGB}{2,154,223}
\definecolor{graphqlpink}{RGB}{236,29,151}
\usepackage{multirow}
\usepackage{listings}
\lstdefinestyle{graphql}{
    language={java},basicstyle=\ttfamily\scriptsize, 
    morekeywords={type, Query, Mutation, schema, query, enum},
    keywordstyle=\bfseries\color{graphqlpink},
    emph={ID, String, Boolean, Float, Int},
    emphstyle={\color{pblue}},
    stringstyle=\bfseries\color{teal}
}

\lstdefinestyle{resolver}{
    basicstyle=\ttfamily\scriptsize, 
    morekeywords={public, return, int, new, function, foreach, as, \$this, array\_column, if}
}

\lstdefinestyle{tests}{
    basicstyle=\ttfamily\scriptsize, 
    morekeywords={public, return, int, new, function, foreach, as, \$this, use, namespace, php, array\_column, if},
    breaklines=true,
    breakatwhitespace=true,
    commentstyle=\bfseries\itshape\color{Fuchsia}
}

\lstdefinestyle{query-entry}{
    basicstyle=\ttfamily\scriptsize, 
    stringstyle=\bfseries\color{pblue}
}

\newcommand{\lstbg}[3][0pt]{{\fboxsep#1\colorbox{#2}{\strut #3}}}
\definecolor{codegreen}{rgb}{0,0.6,0}
\lstdefinelanguage{diff}{
    basicstyle=\ttfamily\scriptsize,
	morecomment=[f][\color{red}]{---}, 
	morecomment=[f][\color{codegreen}]{+++},
	morecomment=[f][\lstbg{red!20}]{-\ },
	morecomment=[f][\lstbg{green!20}]{\ +\ },
	morecomment=[f][\color{blue}]{@@},}
	
\makeatletter
\lst@AddToHook{OnEmptyLine}{\addtocounter{lstnumber}{-1}}
\makeatother

\usepackage[framemethod=tikz]{mdframed}
\mdfdefinestyle{mpdframe}{
    frametitlebackgroundcolor   =RoyalPurple!15,
    frametitlerule              =true,
    roundcorner                 =1pt,
    middlelinewidth             =1pt,
    innermargin                 =0.1cm,
    outermargin                 =0.1cm,
    innerleftmargin             =0.1cm,
    innerrightmargin            =0.1cm,
    innertopmargin              =0.1cm,
    innerbottommargin           =0.1cm,
    linecolor                   =RoyalPurple
}

\def\BibTeX{{\rm B\kern-.05em{\sc i\kern-.025em b}\kern-.08em
    T\kern-.1667em\lower.7ex\hbox{E}\kern-.125emX}}

\newcommand{\autographql}{AutoGraphQL\xspace}
\newcommand{\frontapp}{Frontapp\xspace}
\newcommand{\saleor}{Saleor\xspace}
\newcommand{\saleorschemacovgenerated}{26.9\%\xspace}
\newcommand{\frontappschemacovgenerated}{48.7\%\xspace}

\author[$\dag$]{Louise Zetterlund}
\author[*]{Deepika Tiwari}
\author[*]{Martin Monperrus}
\author[*]{Benoit Baudry}
\affil[$\dag$]{Redeye AB, Sweden}
\affil[*]{KTH Royal Institute of Technology, Sweden}

\begin{document}

\title{Harvesting Production GraphQL Queries\\ to Detect Schema Faults}


\maketitle


\begin{abstract}
GraphQL is a new paradigm to design web APIs. Despite its growing popularity, there are few techniques to verify the implementation of a GraphQL API. 
We present a new testing approach based on GraphQL queries that are logged while users interact with an application in production.
Our core motivation is that production queries capture real usages of the application, and are known to trigger behavior that may not be tested by developers.
For each logged query, a test is generated to assert the validity of the GraphQL response with respect to the schema.
We implement our approach in a tool called \autographql, and evaluate it on two real-world case studies that are diverse in their domain and technology stack: an open-source e-commerce application implemented in Python called Saleor, and an industrial case study which is a PHP-based finance website called Frontapp. \autographql successfully generates test cases for the two applications. The generated tests cover \saleorschemacovgenerated of the Saleor schema, including parts of the API not exercised by the original test suite, as well as \frontappschemacovgenerated of the Frontapp schema, detecting 8 schema faults, thanks to production queries.

\end{abstract}


\section{Introduction}\label{sec:introduction}

Web APIs consist of programmable endpoints to interact with software systems. They can be implemented in different ways, including the well known REST \cite{10.5555/932295} and SOAP \cite{davis2005comparative, 991449} paradigms. 
GraphQL is a new way to define web APIs invented by Facebook in 2015 \cite{graphql_fb}. A GraphQL API implementation consists of a schema that specifies the data structures and operations exposed by the API, as well as a server that implements the logic to handle API requests, resolving them into actual data. The clients of a  GraphQL API, typically a browser or an app, send requests that specify the data they want to retrieve.
The performance \cite{10.1145/3368089.3409670, seabra2019rest} and flexibility \cite{9101226} of GraphQL have contributed to its rapid adoption in the industry~\cite{brito2019migrating, vadlamani2021can}.

While GraphQL offers significant benefits to develop web APIs, correctly implementing it remains a challenge. Specifically, a bug in the server may cause a GraphQL query to be resolved into data that is incompatible with the properties defined in the schema. We designate this kind of fault as a schema fault. 
Let us consider the example of \emph{Saleor}, an e-commerce platform that exposes a GraphQL API. A user reported an error when trying to create a new product without assigning it to a category\footnote{\url{https://github.com/mirumee/saleor/issues/5589}}. This issue was identified as a schema fault and fixed by the maintainers, because the API implementation contradicted the \emph{Saleor} GraphQL schema.
Schema faults are the focus of testing techniques for other systems specified using schemas, such as databases \cite{alsharif2018domino, mcminn2016schemaanalyst}, or OpenAPI REST APIs \cite{corradini2021empirical, karlsson2020quickrest, 8536162}.
However, there has been little work to detect schema faults in GraphQL APIs. Only one approach,  by Karlsson \textit{et al.} \cite{9463000} targets them, by generating GraphQL queries randomly and using them as inputs in property-based tests with the goal of exercising more of the GraphQL schema.

In this work, we propose to harvest GraphQL queries from an application in production, and use them as inputs for test generation. Our motivation is that test cases generated from production GraphQL queries assess the behavior of the application with respect to real API usages. Moreover, it has been shown that production data can lead to valuable test cases that invoke behavior untested by developer-written tests \cite{9526340, 7927986}. 
We implement our technique in a tool called \autographql, which operates in two phases. The first phase involves monitoring an application in production and logging every unique GraphQL query. In the second phase, \autographql generates one test for each query logged in the monitoring phase.
Each generated test includes the required oracles  to assess whether the format of the response is consistent with the GraphQL schema.

We evaluate \autographql on one open-source and one closed-source case study.
Our open-source case study is an e-commerce platform called \emph{Saleor}.
Our closed-source industrial case study, \emph{Frontapp}, is the primary website of Redeye AB, an equity research and investment banking company based in Stockholm, Sweden.
\autographql harvests $334$ and $24,049$ unique GraphQL queries in  production, for Saleor and Frontapp, respectively.
Our tool successfully generates one test for each logged query. The generated tests exercise \saleorschemacovgenerated of the GraphQL schema in Saleor, including parts of the schema not exercised by the original, developer-written test suite.
The tests generated by \autographql for Frontapp exercise \frontappschemacovgenerated of the schema and detect $8$ schema faults.

Our evaluation of \autographql with two diverse case studies demonstrates that it can successfully generate tests with GraphQL queries harvested from production. The generated tests complement developer-written tests by triggering untested behavior, and are able to discover schema faults in the implementation of the GraphQL API.
To sum up, our contributions are as follows:
\begin{itemize}
    \item A novel technique to harvest GraphQL queries from production and use them to generate test cases with oracles tailored for the detection of schema faults
    \item The evaluation of our technique with one industrial and one open-source case study, which demonstrates that the generated tests trigger untested behavior, and discover schema faults
    \item An open-source implementation of our methodology in a tool called \autographql, as well as a publicly available dataset for reproducibility at \url{https://github.com/castor-software/autographql/}
\end{itemize}
The rest of this paper is organized as follows: \autoref{sec:background} discusses GraphQL, \autoref{sec:autographql} introduces \autographql, and \autoref{sec:methodology} describes the methodology we use to evaluate it. We present the results from this evaluation in \autoref{sec:results}, and discuss some aspects of \autographql in \autoref{sec:discussion}. \autoref{sec:related-work} includes related work and \autoref{sec:conclusion} concludes the paper.

\section{Background}\label{sec:background}
This section introduces GraphQL, a specification for web APIs, as well as its implementation.

\subsection{GraphQL APIs}\label{sec:graphql}

GraphQL is a new paradigm to build web APIs.
A GraphQL API consists of one schema that defines the data structures that are available through the API, and a set of requests, or `queries', that can be made against the schema. The implementation of the API is composed of so-called  `resolvers' which map the information requested by the queries to actual data from the underlying database or storage of the application. 

GraphQL requests are typically triggered from clients such as the frontend of an application, and are handled by a server at the backend. Unlike REST APIs \cite{9101226, erlandsson2020performance}, the requests are not centered around resources \cite{10.5555/932295}. Instead, they are structured around operations. There are two types of requests in GraphQL: queries and mutations, defined as follows.
A request that only fetches data, such as the details of a product on a website, is called a \texttt{Query}; a request that changes, or ``mutates'' data, such as adding or updating a shipping address, is called a \texttt{Mutation}. Both kinds of requests are sent to a GraphQL endpoint exposed by the application backend, where they are resolved. The corresponding responses are sent back to the frontend, typically as JSON. 

\subsection{GraphQL Schemas}\label{sec:graphql-schema}

A GraphQL schema serves as a contract between the frontend and the backend of the application \cite{wittern2019empirical}.
A schema is specified with the strongly-typed Schema Definition Language (SDL), which is defined in the GraphQL specification\footnote{\url{https://spec.graphql.org}}. It includes declarations of object, enum, interface, and union types.
The types themselves are composed of fields that define their properties. These fields may be a scalar, such as \texttt{Int}, \texttt{String}, or \texttt{ID}, other object types defined in the schema, or an array thereof. 
An exclamation mark \texttt{!} represents non-nullable fields. In addition to the type declarations, the schema also defines the \texttt{Query} and \texttt{Mutation} operations that can be performed on them.

\import{listings/}{schema.tex}

\autoref{lst:graphql-schema} is an excerpt from the GraphQL schema of Frontapp, the primary website of a company called Redeye AB\footnote{\url{https://www.redeye.se}}, employing one of the authors.
The schema defines an \texttt{interface} called \texttt{Node} which has a non-nullable \texttt{id} of type \texttt{ID}. The schema also defines two object types. The \texttt{Video} type represents a video published on Frontapp. It implements \texttt{Node} and it has non-nullable \texttt{String} fields that specify its \texttt{title} and its \texttt{url}. The field \texttt{videoType} expresses the kind of a video as one of the values enlisted in the enumeration type \texttt{VideoTypeEnum}. A video may also have a \texttt{teaser} of type \texttt{Teaser}, which itself has non-nullable \texttt{String} fields for its \texttt{title} and \texttt{url}, and a nullable \texttt{subTitle}. A teaser also has a \texttt{duration} expressed as a \texttt{Float}. The \texttt{Boolean} field \texttt{publishedOnSite} determines if the teaser has been published on Frontapp.

\texttt{Query} is a special GraphQL type that defines the entry-points of all GraphQL queries that fetch data. 
This excerpt of the Frontapp schema defines two entry-points, \texttt{video} and \texttt{teasers}. A \texttt{Video} object can be fetched through a non-nullable \texttt{id} argument, through the \texttt{video} entry-point. The \texttt{teasers} entry-point returns a list of the first \emph{n} \texttt{Teaser}-type objects, based on the value of \emph{n} provided as the non-nullable \texttt{Int} argument to the variable \texttt{first}.

GraphQL allows the requesting entity to explicitly specify, in a single declarative GraphQL query \cite{cederlund2016performance}, the data or fields required in the response. 
\autoref{lst:graphql-query} shows a query made against the schema defined in \autoref{lst:graphql-schema}. The query is given an explicit name, called its \emph{operation name}, which is \texttt{GetTeasers}. This query is generated by the interactions of end-users as they browse through the videos published on the Frontapp website.
This interaction would trigger the \texttt{teasers} entry-point and fetch a list of objects of type \texttt{Teaser}. The query requests for the \texttt{title}, \texttt{subTitle}, and \texttt{url} of the first \texttt{2} teasers. The meta-field \texttt{\_\_typename}, wherever used in a query, specifies the type of the object at that point in the query.

\begin{lstlisting}[style=graphql, label={lst:graphql-query}, caption={A production GraphQL query made against the schema defined in \autoref{lst:graphql-schema}}, float, numbers=none]
query GetTeasers {
  teasers(first: 2) {
    title
    subTitle
    url
    __typename
  }
}
\end{lstlisting}

\subsection{GraphQL Resolvers}

Resolvers are functions to map each field requested in incoming queries with actual data in the application. The resolvers are not written in GraphQL, they may be implemented in any programming language supported by the underlying GraphQL engine, including Java, JavaScript, PHP, Python, and others\footnote{\url{https://graphql.org/code/\#language-support}}. Therefore, GraphQL API implementations can evolve \cite{9101226} while providing stable APIs. 

\autoref{lst:graphql-resolver} shows a resolver for Frontapp, written in PHP. The resolver fetches the first \emph{n} \texttt{Teaser} objects, from the \texttt{teaserRepository}, which is the component that interacts with the Frontapp database. The resolver then prepares the response, with values for all the fields requested by the query in \autoref{lst:graphql-query} fetched from the database. The response is a list of teaser objects with their \texttt{title}, \texttt{subTitle}, and \texttt{url}. We present the response in \autoref{lst:graphql-response}. It contains only the fields explicitly requested in the query, for the \texttt{first} $2$ \texttt{teasers}, including their \texttt{title}, \texttt{subtitle}, and \texttt{url}, and with their \texttt{\_\_typename} being \texttt{Teaser}.


\begin{lstlisting}[style=graphql,label={lst:graphql-response}, caption={The response for the query in \autoref{lst:graphql-query}}, float, numbers=none]
{
  "data": {
    "teasers": [
      { 
        "title": "Finance 101",
        "subTitle": "The basics of finance",
        "url": "https://youtu.be/dQw4w9WgXcQ",
        "__typename": "Teaser"
      },
      { 
        "title": "Development 101",
        "subTitle": null,
        "url": "https://youtu.be/jNQXAC9IVRw",
        "__typename": "Teaser"
      }
    ]
  }
}
\end{lstlisting}

\section{AutoGraphQL}\label{sec:autographql}

This section describes \autographql, a tool that automatically generates tests for the GraphQL backend of an application. We first discuss schema faults, which are the targets for the tests generated by \autographql. Then, we present an overview of \autographql, and describe the phases in which it operates. We conclude this section by discussing the implementation of the tool.

\subsection{Schema Faults}\label{sec:fault-model}

In this work, we aim at detecting faults in the implementation of the GraphQL resolvers, leading them to return data with a format that does not conform to the schema. 
We call these faults ``schema faults''. They occur in GraphQL APIs when valid queries get resolved into invalid responses, as a result of incorrect mapping between the fields requested and the actual data storage in the application. This response may be sent to the client without the error being explicitly identified as such (say HTTP 5xx or a JSON error object). Such invalid responses basically break the interface contract between the server and the client as specified in the schema.
This kind of fault is common, for example, a user of the e-commerce platform Saleor reported an issue\footnote{\url{https://github.com/mirumee/saleor/issues/6750}}, confirmed by a developer, due to a schema fault. 

In order to detect schema faults, \autographql automatically generates test oracles, derived from the schema. These test oracles determine that the data returned by the API is well-formed, with respect to the schema. For example, the \texttt{Boolean} field \texttt{publishedOnSite} for a \texttt{Teaser} is defined as nullable in the schema in \autoref{lst:graphql-schema}. If the condition on line $9$ is introduced in the resolver presented in \autoref{lst:graphql-resolver}, the \texttt{url} of the teaser object will be resolved to null if \texttt{publishedOnSite} is false or null. This contradicts the schema which specifies that the \texttt{url} of a \texttt{teaser} cannot be null, and is therefore a bug in the implementation of the resolver. This kind of fault would be detected by \autographql.

\begin{lstlisting}[language=diff, style=resolver, label={lst:graphql-resolver}, caption={A resolver that fetches a list of the first \emph{n} \texttt{teasers}, with a schema fault that has just been introduced}, float]
@@ -0 +2 @@
public function resolveTeasers(int $first) {
  $teasers = $this->teaserRepository->findMatching($first);
  $data = [];
  foreach ($teasers as $teaser) {
    $newTeaser = new \stdClass();
    $newTeaser->title = $teaser->getTitle();
    $newTeaser->subTitle = $teaser->getSubTitle();
 +  if ($teaser->isPublishedOnSite()) {
      $newTeaser->url = $teaser->getUrl();
 +  }
    $data[] = ["teaser" => $newTeaser];
  }
  return (array_column($data, "teaser"));
}
\end{lstlisting}

\subsection{Overview of AutoGraphQL}\label{sec:autographql-overview}

\begin{figure}
\centering
\includegraphics[width=\columnwidth]{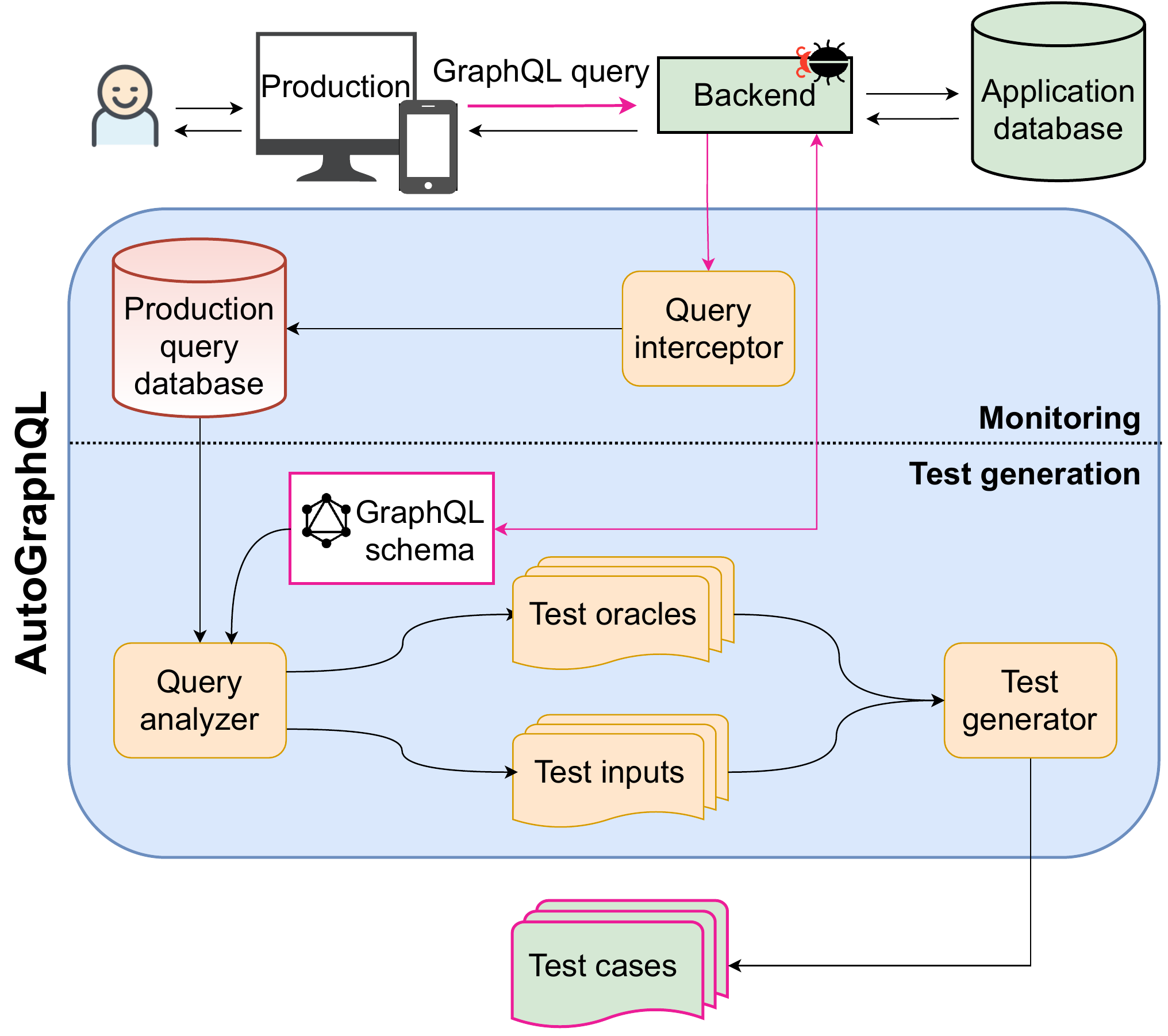}
\caption{Overview of \autographql}
\label{fig:pipeline}
\end{figure}

\autographql generates tests that (i) exercise the GraphQL API implementation, and  (ii) assess that the data returned as a resolution to a GraphQL query conforms to the schema. \autographql operates in two phases, illustrated in \autoref{fig:pipeline}. 
The first phase consists in monitoring the application in production, in order to collect the queries that are performed by users, as well as their arguments. This data collection process is performed for a given amount of time, decided by developers. 
The second phase of \autographql is triggered by developers and consists in analyzing the data that was observed in production to turn them into test cases. \autographql automatically extracts test inputs from the monitored data, as well as the corresponding oracles from the GraphQL schema. The tests generated by \autographql use a single GraphQL query as the test input and verify the format of the response to the query. The following two subsections discuss the details of each phase of test generation with \autographql.


\subsection{Monitoring in Production}\label{sec:production-monitoring}

\begin{table*}
\centering
\caption{The oracles generated by \autographql depending on the response}
\label{tab:generated-oracles}
\begin{tabularx}{\textwidth}{rlX}
\toprule
\textbf{CATEGORY} & \textbf{ORACLE} & \textbf{IMPLEMENTATION AS PHPUnit ASSERTION} \\
\midrule
\multirow{3}{*}{Format} & HTTP status code of the response is $200$ & \texttt{assertEquals(200, ...)}\\
 & Response to query is well-formed, valid JSON & \texttt{json\_decode} doesn't throw exceptions; \texttt{assertIsArray(...)}\\
 & Response does not contain a JSON error object & \texttt{assertArrayNotHasKey('errors', ...)}\\
\midrule
\multirow{4}{*}{Schema} & Response contains all requested fields & \texttt{assertArrayHasKey(...)}\\
 & Correct kind of each element in response & \texttt{assertIsArray(...)} \texttt{assertContains(...)} \texttt{assertEquals(...)}\\
 & Correct type of each element in response & \texttt{assertIsString(...)} \texttt{assertIsBool(...)} \texttt{assertIsNumeric(...)} \texttt{assertIsInt(...)} \texttt{assertEquals(...)} \\
 & Nullability-contract of each field in response & \texttt{assertNotNull(...)}\\ \bottomrule
\end{tabularx}
\end{table*}

Monitoring GraphQL queries in production constitutes the first phase of \autographql, and is illustrated in the top-half of \autoref{fig:pipeline}.
\autographql's \emph{query interceptor} monitors the requests sent from the frontend of an application to its backend. It intercepts incoming GraphQL query requests, and the arguments with which they were invoked, and logs them into a database. It also aggregates metadata about these queries, including the number of times a specific query request was invoked or when it was last invoked.
The output from this phase is a database of GraphQL queries logged from production.

\autoref{lst:query-entry} presents an example of a logged query. The keys \texttt{query} and \texttt{variables} represent the actual query executed as well as the argument passed to it, respectively. In this case, the query is the same as in \autoref{lst:graphql-query}, and the value of \texttt{2} is passed as argument for \texttt{first}, to fetch the first 2 \texttt{teasers}. The entry also includes the operation name for the query (\texttt{operation\_name}), which is \texttt{GetTeasers}. In order to gather statistics about the queries triggered in production, we also save timestamps for when they are first logged (\texttt{created\_at}) and when they are logged most recently (\texttt{updated\_at}). Moreover, during our experiments, we observe that a query may frequently be invoked with the same arguments. We create one entry for each unique combination of query and arguments, and record the frequency of its occurrence with the \texttt{times\_called} field. The value of $301,016$ means that this combination of query and argument occurred as many times, in production, during the course of our experiment.

\import{listings/}{query-entry.tex}

\subsection{Test Generation}\label{sec:test-generation}
The second phase of \autographql, presented in the bottom-half of \autoref{fig:pipeline}, is triggered by developers whenever they want to generate tests after a period of monitoring. It involves automatically fetching the GraphQL schema of the application from the configured GraphQL endpoint, and using this schema in conjunction with the queries logged in the monitoring phase, in order to produce the inputs and oracles for the generated tests. The output from this phase is the test suite generated from the logged production queries. We now discuss the two components that are responsible for analyzing the logged queries and using them to generate tests.

\subsubsection{Query Analyzer}
The goal of the query analyzer is to use the GraphQL schema of an application and the queries logged in the query database to generate test inputs and their corresponding oracles.
Given a logged GraphQL query, such as the one in \autoref{lst:query-entry}, the analyzer extracts the test input, which is the combination of \texttt{query} and its associated \texttt{variables}, as well as the \texttt{operation\_name} given to the query.
Next, the query analyzer produces two sets of oracles. We summarize them in \autoref{tab:generated-oracles}, together with their implementation as assertions in the PHPUnit framework\footnote{\url{https://phpunit.de}}.
The first set of oracles verify the format of the response, specifically i) its HTTP status code, ii) the validity of the JSON text, and iii) that it does not contain a JSON error object because of an invalid request.
The second set of oracles are specific to the schema and i) verify that the response contains all the data requested by the query, and ii) map each object and field requested by the query with the properties defined for it in the schema. These properties include its type, its kind, i.e., whether it is an object, enum, list, or interface, and its nullability.

For example, a subset of the oracles produced for the query in \autoref{lst:query-entry} is that the response would contain a field called \texttt{title} (\texttt{assertArrayHasKey('title', ...)}), which is a non-null (\texttt{assertNotNull(...)}) \texttt{String} type object (\texttt{assertIsString(...)}).

\subsubsection{Test Generator}\label{sec:test-generator}
The test generator uses the test input and test oracles produced by the query analyzer in order to generate a valid and executable test case that verifies the implementation of the GraphQL API for the application. The output of the test generator is one test case for each logged query. By default, \autographql generates tests in PHP, using the PHPUnit framework as test driver.

\import{listings/}{generated-test.tex}

\autoref{lst:generated-test} shows the test generated for the query in \autoref{lst:query-entry}. The test fetches the response to the query (lines $10$ to $23$) by sending it as an HTTP request to the GraphQL endpoint of an application (lines $25$, $26$). After verifying its HTTP status code (line $28$), the test decodes the response and verifies that it is well-formed JSON (lines $30$, $31$). Next, the assertion on line $33$ ensures that the response does not contain an error due to an invalid query, due to the query trying to fetch data that does not exist, or an exception being raised during query resolution. Lines $36$ to $51$ contain the assertions produced by the test generator using the oracles derived from the schema. We use the assertion available in PHPUnit to check the validity of collections: \texttt{assertIsArray} verifies that \texttt{teasers} is a list. Next, \texttt{for} each of the items within \texttt{teasers}, \texttt{assertEquals} verifies that its \texttt{\_\_typename} is \texttt{Teaser}.
\texttt{assertArrayHasKey} assertions verify that each of the \texttt{teaser} objects in the list has a \texttt{title}, a \texttt{subTitle}, and a \texttt{url}, as requested by the query. The assertions for the \texttt{title} and the \texttt{url} of a \texttt{Teaser} are \texttt{assertNotNull} since they are defined as non-nullable, per the schema in \autoref{lst:graphql-query}. Moreover, \texttt{assertIsString} verifies that the \texttt{title} and \texttt{url}, and \texttt{if} present, the \texttt{subTitle} of a \texttt{teaser} are all strings.
When this test is executed, assuming the server returns  the response shown in \autoref{lst:graphql-response}, $22$ assertions are evaluated in total.
A failure in any of the assertions in a generated test causes the test to fail, which could be indicative of a schema fault. On the other hand, a generated test that passes can serve as a regression test.

\subsection{Challenges}\label{sec:challenges}
We now discuss the challenges of test generation with \autographql.

\emph{Query Interception}:
In order for the query interceptor of \autographql to monitor and log incoming GraphQL queries, it must be tailored to fit the technology stack of an application. For example, the configuration of the query interceptor used by an application with a backend implemented in Python would differ from the one implemented in Ruby. This is a potentially significant engineering effort.

\emph{Testing Database}:
The execution of the generated test suite requires a running application server with a testing database. This is typically provided by a staging environment.
The state of the staging database may have an impact on the execution of the generated tests. Thus an important engineering challenge is to be able to re-initialize a clean staging database before running the \autographql tests.


\subsection{Implementation}
\autographql is implemented in Python. The query analyzer uses the GraphQLParser of graphql-py\footnote{\url{https://github.com/ivelum/graphql-py}} to map the elements of a query with the schema and produce test oracles. By default, \autographql populates a template with the test input and oracles in order to produce tests in the PHPUnit framework. This allows the properties of each node to be expressed as PHPUnit assertions, which serve as oracles in the generated test. Jinja2\footnote{\url{https://jinja.palletsprojects.com/en/2.11.x/}} is the templating language used to render the assertions into PHPUnit test files. We choose PHPUnit for the generated tests, since PHP is a popular server-side language for the implementation of web APIs \cite{imtiaz2021automated}. \autographql can generate tests for applications that do not use PHP, or even be extended to support any testing framework, since the generated tests only interact with the HTTP GraphQL endpoint of the application.

\section{Evaluation Methodology}\label{sec:methodology}

This section describes our two real-world case studies, \emph{Saleor} and \emph{Frontapp}. We also present our experimental setup, including the configuration of \autographql with the two case studies, their production workloads, as well as the metrics used to evaluate the effectiveness of \autographql in generating tests for them.

\subsection{Case Studies}\label{sec:case-studies}
We use one open-source and one industrial project as case studies in order to evaluate the effectiveness of \autographql. We describe these projects and some relevant metrics below.

\begin{table}
\centering
\caption{Case studies for the evaluation of AutoGraphQL}
\label{tab:graphql-projects}
\begin{tabular}{lrrrrrrr}
\toprule
\textbf{PROJECT}& \textbf{LOC}& \textbf{COMMITS} & \textbf{LANGUAGE} & \textbf{DOMAIN}\\
\midrule
Saleor & 691K & 17.4K & Python & E-commerce \\
Frontapp & 154K & 7K & PHP & Finance \\
\bottomrule
\end{tabular}
\end{table}

\subsubsection{Saleor}
Saleor is a widely-used, open-source e-commerce platform, maintained by more than $170$ contributors. The project has more than $13K$ stars on GitHub. It is a well-documented and mature project, that can be deployed with Docker. Saleor is crafted with modern technologies, such as Django\footnote{\url{https://www.djangoproject.com/}}, PostgreSQL, Redis, React, and TypeScript. It offers both a storefront for customers to browse through a catalog of products and make purchases, as well as a dashboard for administrators to manage products, users, and orders. Incoming requests from both of these frontends are handled by a GraphQL server implemented as part of the core component of Saleor. The test suite of Saleor contains automated tests for both the backend and the two frontend components. We choose Saleor as a case study in order to have reproducible experiments with an open-source project, and to demonstrate the versatility of AutoGraphQL in generating tests that target the GraphQL implementation of an application, regardless of the underlying backend technology. 

\autoref{tab:graphql-projects} summarizes the key characteristics of the case studies: the number of lines of code and of commits, the language implementing the GraphQL API and the domain of the case study.
We use the latest stable release of Saleor, version $2.11$, for our experiments. As mentioned in the table, this version contains $691K$ lines of code  and the backend is in Python.


\subsubsection{Frontapp}
Frontapp is the primary website of our industrial partner, Redeye AB. Frontapp contains articles, financial analyses, tools, and video streams of events hosted by Redeye. The site is implemented in Symfony\footnote{\url{https://symfony.com/}}, a web application framework for PHP projects, and in JavaScript. 
Its GraphQL API is connected to multiple data sources and receives approximately $64K$ requests daily. 
Frontapp has been in production for more than $7$ years. There is no automated test for the application, and it is tested manually by the QA team before major versions are released. 

As presented in \autoref{tab:graphql-projects}, more than $20$ Redeye developers have contributed about $7K$ commits to Frontapp. The application  contains nearly $154K$ lines of code (LOC), as measured on February 09, 2021, and the GraphQL API is implemented in PHP.

\subsection{Experiments}\label{sec:experiments}
This section describes the experimental protocol followed and the metrics used to evaluate \autographql with Frontapp and Saleor.

\subsubsection{Query Interceptor Configuration}
As described in \autoref{sec:challenges}, the query interceptor of \autographql is specific to a given software stack.
In order to conduct experiments with Saleor, we extend it with an agent, implemented as a GraphQL middleware \cite{graphql_middleware}. This agent allows the query interceptor of \autographql to access queries that arrive at the GraphQL endpoint of Saleor, \texttt{/graphql/}, and log them into the query database.
For our experiments with Frontapp, we configure a PHP event listener that triggers the query interceptor, which then logs all queries arriving on the GraphQL endpoint, which is also \texttt{/graphql/}.


\subsubsection{Production Workloads}\label{sec:production-workloads}
\autographql generates tests for queries that are triggered by user actions as part of interactions in production, during the monitoring phase of \autographql.
We define such a production workload for each case study, as follows.

For our experiments with Saleor, we deploy the e-commerce application in a local server in our laboratory. In order to produce a realistic production workload, one of the authors interacted with the components on the frontend to perform typical operations related to e-commerce websites, such as browsing through the catalog of products, searching for specific products from the search bar, viewing the web-page for a product, and making orders.
Additionally, the author performed administrative actions such as fetching the list of registered customers and orders, or searching for a specific customer or order. The experiment was carried out over the duration of nearly $3$ hours.

The production workload for Frontapp consists of the interactions of actual end-users with the system. We do not ask the end-users to perform any specific operations, and simply log the queries that are generated as they browse through the website, reading articles or using its search feature, etc. We log these queries for a period of $33$ days.

\subsubsection{Metrics for Evaluation}\label{sec:metrics}

In order to gauge the effectiveness of the tests generated by \autographql, we adopt the concept of schema coverage introduced by Karlsson \textit{et al.} \cite{9463000}. Compared to traditional code coverage, schema coverage is a more relevant metric to assess the tests generated by \autographql since they are intended to directly target the GraphQL schema.

\emph{Schema Coverage}:
We consider a GraphQL schema to be composed of a set of tuples of the form \emph{\{Object\textsubscript{o}, Field\textsubscript{n}\}}, by combining all \emph{Object} \emph{o}, defined as a \texttt{type} or an \texttt{interface} in the schema, with each of its \emph{n} fields.
A query is said to reach a tuple \emph{\{o, f\}} if it requests for a field \emph{f} defined in the object \emph{o}. This is determined statically by analyzing the Abstract Syntax Tree of the query.
The schema coverage (SCHEMA\_COV) of a test, or the test suite, generated by \autographql is then defined as the number of tuples in the schema that are reached by the query, or set of queries, invoked by the test(s) (COVERED\_TUPLES), divided by the total number of tuples in the schema (SCHEMA\_TUPLES). This is presented in \autoref{eq:schema-cov}.

\begin{equation}\label{eq:schema-cov}
    \textsc{SCHEMA\_COV} = \frac{\textsc{COVERED\_TUPLES}}{\textsc{SCHEMA\_TUPLES}}
\end{equation}
\vskip0.5\baselineskip


A schema coverage of $0\%$ means that the test suite of a project does not invoke any queries that cover a tuple in the schema. On the other hand, a schema coverage of $100\%$ would imply that all the tuples in the schema are covered by the test suite.
For example, the query in \autoref{lst:graphql-query} covers $4$ tuples, \emph{\{Query, teasers\}, \{Teaser, title\}, \{Teaser, subTitle\}, and \{Teaser, url\}}, of the $13$ tuples of the schema in \autoref{lst:graphql-schema}. The schema coverage of the corresponding test in \autoref{lst:generated-test} is therefore $30.8\%$.

\emph{Metrics}:
We collect and report the following metrics for Frontapp and Saleor, based on their schema, logged production queries, test generation, and test execution:

\begin{itemize}
    \item [1] TYPES is the number of types defined in the schema.
    \item [2] ENTRY\_POINTS is the number of entry-points defined in the \texttt{Query} type of the schema.
    \item [3] UNIQUE\_QUERIES is the number of unique combinations of queries and arguments logged during the experiment, and consequently, the number of tests generated.
    \item [4] ASSERTIONS\_EVALUATED is the total number of assertions evaluated on executing the generated tests.
    \item [5] PASSING is the number of generated tests that pass.
    \item [6] FAILING is the number of generated tests that do not pass.
    \item [7] SCHEMA\_FAULTS is the number of bugs found by the generated tests.
    \item [8] SCHEMA\_TUPLES is the number of tuples obtained from the schema.
    \item [9] COVERED\_TUPLES is the number of tuples of the schema reached by the generated tests.
    \item [10] SCHEMA\_COV\_GENERATED is the schema coverage achieved with the test suite generated by \autographql, per \autoref{eq:schema-cov}.
\end{itemize}

All metrics are integer quantities, except for SCHEMA\_COV\_GENERATED which is expressed as a percentage.

\section{Evaluation Results}\label{sec:results}
This section presents the results obtained during our experiments with the two case studies. We summarize the results for all metrics introduced in \autoref{sec:metrics} in \autoref{tab:autographql-results}.

\begin{table}
\centering
\caption{Results from the evaluation of \autographql on the two case studies}
\label{tab:autographql-results}
\begin{adjustbox}{width=\textwidth/2}
\begin{tabular}{rlrr}
\toprule
\textbf{\#} & \textbf{Metric} & \textbf{Saleor} & \textbf{Frontapp} \\
\midrule
1 & TYPES & 460 & 92 \\
\midrule
2 & ENTRY\_POINTS & 69 & 23 \\
\midrule
3 & UNIQUE\_QUERIES & 334 & 24,049 \\
\midrule
4 & ASSERTIONS\_EVALUATED & 88,668 & 8,727,519 \\
\midrule
5 & PASSING & 334 & 23,892 \\
\midrule
6 & FAILING & 0 & 157 \\
\midrule
7 & SCHEMA\_FAULTS & 0 & 8 \\
\midrule
8 & SCHEMA\_TUPLES & 1884 & 875 \\
\midrule
9 & COVERED\_TUPLES & 506 & 426 \\
\midrule
10 & SCHEMA\_COV\_GENERATED & 26.9\% & 48.7\% \\
\midrule
\end{tabular}
\end{adjustbox}
\end{table}

\subsection{Case Study 1: Saleor}\label{sec:results-saleor}
As presented in \autoref{tab:autographql-results}, the GraphQL schema of Saleor defines $460$ TYPES and $69$ query ENTRY\_POINTS. Based on the production workload defined in \autoref{sec:production-workloads}, \autographql logs $334$ UNIQUE\_QUERIES, and generates one test for each of them. We successfully execute all of these tests.
The generated test suite triggers $43$ of the $69$ query entry-points in the schema. These tests cover $506$ tuples (COVERED\_TUPLES) out of the $1884$ SCHEMA\_TUPLES in Saleor. This results in a value of \saleorschemacovgenerated for SCHEMA\_COV\_GENERATED.

The execution of the $334$ test cases triggers the evaluation of $88,668$ assertions (ASSERTIONS\_EVALUATED). The difference between the number of tests and the number of assertions evaluated within the tests is because some assertions are made inside loops to verify the properties of elements that are lists, as illustrated on line $39$ of \autoref{lst:generated-test}.
Each of the $88,668$ assertions verifies one expected property about the returned data, per the GraphQL schema.
None of the $88,668$ assertion evaluations fail, meaning that the generated test cases based on our selected production workload, do not reveal a schema fault in version $2.11$ of Saleor's GraphQL resolvers. This is to be expected given the popularity and maturity of Saleor.

Saleor has a solid test suite written by the developers. Now we want to assess the complementarity of the original tests and the test cases generated by \autographql.
In particular, we consider two metrics:
SCHEMA\_COV\_ORIGINAL is the schema coverage achieved with the GraphQL requests triggered by the original test suite.
DISTINCT\_TUPLES is the number of schema tuples not covered by the original test suite but covered by the test suite generated by \autographql. A non-zero value for DISTINCT\_TUPLES would imply that \autographql is able to generate valuable new tests. 

The original test suite of Saleor includes $5405$ test cases, which trigger $2340$ requests, of which $1227$ are queries and $1113$ are mutations. \autoref{tab:saleor-tuples} shows the number of tuples involved in those tests. 
This original test suite covers $1429$ of the $1884$ tuples, resulting in a value of $75.8\%$ for SCHEMA\_COV\_ORIGINAL. 
The \autographql test suite covers $506$ tuples, including
$483$ tuples covered by the original as well as the generated test suite. Most importantly, the tests generated by \autographql using production queries cover $23$ DISTINCT\_TUPLES in the Saleor schema that are never covered by the original test suite, including one query entry-point. This confirms the findings of Wang \textit{et al.} \cite{7927986} and Tiwari \textit{et al.} \cite{9526340} that in-house tests can miss behavior that is exercised in production. These unique tuples covered by the generated tests complement the existing test cases, and the generated tests would contribute to the prevention of regression bugs in the resolvers that handle these tuples. We also note that $432$ of the $1884$ tuples in the schema are covered neither by the original tests, nor by the generated tests, showing that comprehensive schema testing is hard. We will discuss the possible reasons why parts of the schema are not covered by the generated tests in more detail in \autoref{sec:discussion}. 

\begin{mdframed}[style=mpdframe,frametitle=Highlight from the Saleor experiment]
With the $334$ GraphQL queries harvested during our experiments with Saleor, \autographql generates $334$ test cases. These tests cover \saleorschemacovgenerated of the schema, including $23$ tuples in the schema that have not been covered by the developers in the original test suite. This reveals that the \autographql tests complement the original test suite with respect to the capability of detecting schema faults in GraphQL resolvers.
\end{mdframed}

\begin{table}
\centering
\caption{Coverage of Saleor schema tuples with original and generated tests}
\label{tab:saleor-tuples}
\begin{tabularx}{\columnwidth}{Xr}
\toprule
& \textbf{COVERED\_TUPLES} \\
\midrule
Original test suite & \textbf{1429} / 1884 \\
\autographql-generated test suite & \textbf{506} / 1884 \\
Intersection of \autographql and original test suites & \textbf{483} / 1884 \\
Tuples only covered by \autographql tests \newline (DISTINCT\_TUPLES) & \textbf{23} / 1884 \\
\bottomrule
\end{tabularx}
\end{table}

\subsection{Case Study 2: Frontapp}\label{sec:results-frontapp}
From \autoref{tab:autographql-results}, we see that the schema of Frontapp defines $92$ TYPES and $23$ query ENTRY\_POINTS. The Frontapp schema does not define \texttt{Mutation} operations.
The production workload of Frontapp, described in \autoref{sec:production-workloads},
observed over a monitoring period of $33$ days, results in AutoGraphQL harvesting and storing $24,049$ UNIQUE\_QUERIES. The query most frequently invoked during this period was executed $301,016$ times and is presented in \autoref{lst:query-entry}. Using the logged queries, \autographql generates $24,049$ PHPUnit tests, all of which are successfully executed. Frontapp has $875$ SCHEMA\_TUPLES of which $426$ are covered by the generated tests (COVERED\_TUPLES), resulting in a SCHEMA\_COV\_GENERATED of \frontappschemacovgenerated. The generated tests trigger $19$ of the $23$ entry-points in the schema. The number of ASSERTIONS\_EVALUATED on the execution of the generated tests is $8,727,519$.

Of the $24,049$ generated tests, $157$ fail. The developers at Redeye confirmed that these failures are caused by $8$ distinct SCHEMA\_FAULTS in Frontapp. The difference between the number of failures and the number of schema faults is the result of some test cases triggering the same query entry-points, and therefore, the same bugs. These schema faults are caused due to incorrect assumptions about the properties of objects, causing them to contradict the properties defined in the schema. For example, a nullable variable was sent to a resolver that could not handle a null input, or an element was collected in a non-nullable array without being checked for null first, or a resolver returned a different type than was stated in the schema.

\begin{lstlisting}[language=diff, style=resolver, label={lst:frontapp-bug}, caption={A schema fault discovered in Frontapp, and its resolution}, float]
@@ -0 +3 @@
 + if (is_array($source['authorIds']) &&
 +     count($source['authorIds']) > 0) {
     if (id::isValid($source['authorIds'][0])) {
       $firstAuthor = $this->personRepository->findById(
         id::fromString($source['authorIds'][0])
       );
       if ($firstAuthor && $firstAuthor->getTitle()) {
         $authorTitle = $firstAuthor->getTitle();
       }
     }
 + }
\end{lstlisting}

Let us now look at an example of a schema fault found by a generated test. \autoref{lst:frontapp-bug} shows a bug located within a resolver defined in Frontapp. This bug was caused because the field \texttt{authorIds}, defined as non-nullable in the Frontapp schema, actually had the value of \texttt{null}, causing an exception to be raised (line $4$). It was discovered due to a failing assertion in a test generated by \autographql, specifically the assertion that checks if the response has an \texttt{errors} field. The bug was consequently fixed by Frontapp developers by adding the highlighted check (lines $1$ and $2$) to ensure that \texttt{authorIds} is indeed not null before performing further computations on it.

As mentioned in \autoref{sec:case-studies}, we note that Frontapp does not have automated tests. Thus, the developers at Redeye proved to be interested in the \autographql test cases, which complement their manual QA activities. Furthermore, it is a possibility to push the \autographql tests in a repository with continuous integration, and to run them regularly to identify regressions or new schema faults as the application continues to evolve. 

\begin{mdframed}[style=mpdframe,frametitle=Highlight from the Frontapp experiment]
\autographql harvests $24,049$ GraphQL queries that are triggered due to interactions of Frontapp end-users in production, and generates as many tests. The generated test suite achieves a schema coverage of \frontappschemacovgenerated and discovers $8$ schema faults. Those faults have subsequently been fixed by the developers. This validates the capability of \autographql to automatically generate valuable tests that detect faults, from GraphQL queries observed in production.
\end{mdframed}

\section{Discussion}\label{sec:discussion}
We now reflect on some interesting aspects of \autographql. 

\emph{Test Minimization and Prioritization}:
Test generation with \autographql is systematic in that each unique query harvested from production is used as input for the generation of one test. Thus, the number of harvested queries determines the size of the generated test suite, and consequently its execution time.
For example, the $334$ generated tests for \saleor are executed in $34$ seconds, while the $24,049$ tests generated for \frontapp take $114$ hours to execute. In situations where the execution time is critical, such as in a continuous integration pipeline, it would be useful to minimize the generated test suite, as well as prioritize the execution of the tests \cite{liang2018redefining, lima2020test}.
As described in \autoref{sec:production-monitoring}, the query interceptor of \autographql aggregates meta-data about each query logged during the monitoring phase, including the number of times it was observed, as well as timestamps for when it was first and last invoked. This information may be used by developers to filter a subset of queries to use as inputs for test generation. For example, a developer may generate tests using the queries triggered at least $500$ times within the last $3$ days, and execute these tests in a prioritized fashion, based on criteria such as their schema coverage.

\emph{Mutation Requests}: 
State of the art techniques for GraphQL, including cost analysis \cite{10.1145/3368089.3409670, mavroudeas2021learning}, formal analysis \cite{hartig2018semantics}, and GraphQL test generation \cite{vargas2018deviation} only support query requests, with the exception of \cite{9463000} which provides support for the generation of mutation requests.
Outside the academic literature, \emph{Schemathesis} is an open-source tool that uses the GraphQL schema to generate property-based tests, it also only supports query requests\footnote{\url{https://schemathesis.readthedocs.io/en/stable/graphql.html}}.
However, we note that mutation operations are an equally integral part of GraphQL APIs. 
For example, the \saleor schema defines $222$ mutation entry-points. Thus, there is a clear need for research on mutation requests.
This is an interesting and challenging research endeavour because mutation requests have side-effects on the application database, which may result on breaking tests depending on test ordering and breakage of various assumptions on the application state.

\emph{GraphQL Schema Evolution and Test Generation}:
A GraphQL schema and the implementation of the corresponding resolvers may evolve at a different pace. Some parts of the schema may correspond to functionality that is slated to be deprecated, such as the field in a type that is to be replaced by another field. The schema might also specify elements that are yet to be implemented as part of a future release. For example, the developers at Redeye confirm that the \frontapp schema specifies more elements than those which can be handled by the current resolvers. This is due to the fact that Redeye began the process of migrating \frontapp from a REST API to a GraphQL API in 2019 \cite{brito2019migrating}.
This impacts the schema coverage achieved with the tests generated by \autographql, since there are tuples in the schema that are unreachable by design.
For the same reason, the \autographql tests may become outdated and even break when the schema evolves.
Therefore, an important direction for research in automated test generation for GraphQL is to understand how the tests may be evolved to address the schema evolution \cite{zaidman2008mining}. 
One approach to evolve these tests could involve repairing the generated test suite \cite{hammoudi2016waterfall}.

\section{Related Work}\label{sec:related-work}

This section discusses closely related work in the areas of test generation for web APIs, for data schemas, and based on production.

\subsection{Test Generation for Web APIs}
Currently, only two studies propose automated test generation strategies for GraphQL APIs. Vargas \textit{et al.} \cite{vargas2018deviation} mutate GraphQL queries in existing tests in order to amplify them \cite{DANGLOT2019110398}.
Karlsson \textit{et al.} \cite{9463000} produce randomly-generated queries and arguments based on the GraphQL schema, and use them as inputs in property-based tests.
\autographql differs from these approaches because the inputs for test generation are not existing test queries or randomly generated ones, but queries that are harvested from production.

Several studies propose black-box test generation approaches for REST APIs \cite{corradini2021empirical}. In addition to the HTTP status code of the response, the oracles are often derived from OpenAPI/Swagger specifications describing the API. The parameters used as test inputs may be derived from the API specification \cite{8536162}, or produced randomly \cite{karlsson2020quickrest, atlidakis2019restler}. Recently, deep learning models have been proposed to determine the validity of these inputs \cite{9476896}.
The generated tests can assess the robustness of the API through invalid requests \cite{laranjeiro2021black}, detect regressions across API versions \cite{godefroid2020differential}, verify the data dependencies among sequences of requests \cite{viglianisi2020resttestgen}, or verify the constraints imposed on their parameters \cite{martin2020restest}. Metamorphic relations among requests may also serve as the oracle \cite{segura2017metamorphic}.
EvoMaster \cite{10.1145/3293455} is a search-based, white-box approach to generate tests for RESTful web services. The technique is based on an evolutionary algorithm which rewards code coverage and fault-finding ability, the latter being determined by HTTP status codes.
\autographql is a black-box approach, that is fundamentally different from these test generation techniques because it is tailored to GraphQL APIs, and uses GraphQL schemas and requests monitored in production.

\subsection{Test Generation for Data Schemas}
Traditional databases are also defined with a schema, as is GraphQL. Several studies use database schemas in the context of testing. For example, Khalek and Khurshid \cite{10.1145/1858996.1859063} use SQL schemas to generate SQL queries, test data, and oracles verifying the result of query execution, with the goal of testing database engines. McMinn \textit{et al.} \cite{7816513} and Alsharif \textit{et al.} \cite{alsharif2018domino} propose search-based approaches that use the schema to generate test data for covering database integrity constraints. QAGen by Binnig \textit{et al.} \cite{binnig2007qagen} generates meaningful test inputs based on the schema. XML schemas have also been used to produce XML instances automatically, which may be used as inputs for testing web services \cite{4296715, almendros2015xquery, petrova2015automatic, lee2001generating}.
\autographql relates to this domain, but in a new technological context, that of web APIs and GraphQL: it uses the GraphQL schema of an application to produce oracles in the generated tests that verify the format of the GraphQL responses. 

\subsection{Test Generation Based on Production}
A few studies propose test generation strategies using information obtained from production. For example, Oracle Database Replay \cite{wang2009real} and Snowtrail \cite{yan2018snowtrail} capture production queries made against databases, and replay them in order to detect regressions. Marchetto \textit{et al.} \cite{marchetto2008state} use event logs to generate Selenium tests for web applications. Hammoudi \textit{et al.} \cite{hammoudi2016waterfall} incrementally repair tests for web applications generated from record-replay tools. Tiwari \textit{et al.} \cite{9526340} monitor methods of interest in production in order to generate differential unit tests. Thummalapenta \textit{et al.} \cite{thummalapenta2010dygen} use execution traces to generate parameterized unit tests. ReCrash by Artzi \textit{et al.} \cite{artzi2008recrash} reproduces failures through unit tests generated from runtime observations.
\autographql is the first tool that harvests GraphQL queries from production to use as inputs for the generation of test cases.

\section{Conclusion}\label{sec:conclusion}
GraphQL is a new way to specify web APIs. Though it continues to gain widespread adoption, few studies propose automated test generation strategies that target GraphQL API implementations. This paper introduces \autographql, the first tool that leverages production GraphQL queries to automatically generate test cases. The goal of the generated tests is to detect schema faults through oracles that verify that the response to a query conforms with the GraphQL schema.

We present the evaluation of \autographql on one open-source and one industrial case study, called \saleor and \frontapp: \autographql successfully generates tests for both projects. The tests generated for Saleor exercise \saleorschemacovgenerated of the schema and cover regions in the GraphQL schema that are not covered by its original test suite. The tests generated for Frontapp exercise \frontappschemacovgenerated of the schema and reveal $8$ distinct schema faults.
These experiments demonstrate that \autographql is capable of generating tests for untested behavior, as well as detecting errors that occur in the production environment.

An important future direction for \autographql is to analyze how these tests may be incorporated into a continuous integration pipeline. This would require the generated test suite to be minimized, and for the tests to run in a prioritized fashion. It would also be useful to understand how these tests may be evolved as a result of changes made to the API. 

\bibliographystyle{ieeetr}
\bibliography{main}

\end{document}

%% file: listings/schema.tex
\vskip1.5\baselineskip
\noindent\begin{minipage}[t]{\columnwidth}
\centering
\begin{minipage}{.49\textwidth}
\begin{lstlisting}[style=graphql, numbers=none, showlines=true basicstyle=\ttfamily\scriptsize]
interface Node {
  id: ID!
}

enum VideoTypeEnum {
  ANIMATION
  LIVE_ACTION
  SCREENCAST
}

type Video implements Node {
  id: ID!
  title: String!
  url: String!
  videoType: VideoTypeEnum
  teaser: Teaser
}
\end{lstlisting}
\end{minipage}
\begin{minipage}{.49\textwidth}
\centering
\begin{lstlisting}[style=graphql, numbers=none, showlines=true, basicstyle=\ttfamily\scriptsize]
type Teaser {
  title: String!
  subTitle: String
  publishedOnSite: Boolean
  url: String!
  duration: Float
}

type Query {
  video(id: ID!): Video
  teasers(first: Int!): [Teaser]
}

\end{lstlisting}
\end{minipage}
\captionof{lstlisting}{An excerpt from the GraphQL schema of Frontapp}
\label{lst:graphql-schema}
\end{minipage}
\vskip1.5\baselineskip

%% file: listings/query-entry.tex
\begin{lstlisting}[style=query-entry, caption=A logged GraphQL query from production,  label={lst:query-entry}, float, numbers=none]
{
  "query":
    "query GetTeasers($first: Int!) {
      teasers(first: $first) {
        title
        subTitle
        url
        __typename
    }
  }",
  "variables": {
    "first": 2
  },
  "operation_name": "GetTeasers",
  "created_at": "2021-03-03 08:57:46",
  "updated_at": "2021-05-05 16:55:19",
  "times_called": 301016
}

\end{lstlisting}

%% file: listings/generated-test.tex
\begin{lstlisting}[style=tests, caption={A PHPUnit test generated using the logged query in \autoref{lst:query-entry}, based on the schema in \autoref{lst:graphql-schema}}, label={lst:generated-test}, float]
<?php declare(strict_types=1);

namespace GraphQL;
use PHPUnit\Framework\TestCase;
use Symfony\Bundle\FrameworkBundle\Test\WebTestCase;
use Symfony\Component\HttpFoundation\Request;

class GetTeasersTest extends WebTestCase {
  public function testGraphQL() {
    $client = static::createClient();
    
    /* Use the details from the logged query */
    $query = <<<'JSON'
    {
      "query": "query GetTeasers($first: Int!) {
        teasers(first: $first) {
          title
          subTitle
          url
          __typename
        }
      }",
      "variables": { "first": 2 },
      "operationName": "GetTeasers"
    }
    JSON;
    
    /* Make an HTTP request with the query */
    $client->request('POST', '/graphql/', [], [], ["CONTENT_TYPE" => 'application/json'], $query);
    $response = $client->getResponse();

    /* Verify the HTTP status code of the response */
    $this->assertEquals(200, $response->getStatusCode());
    
    /* Decode the response and verify that it contains valid JSON */
    $responseArray = json_decode($response->getContent(), true);
    $this->assertIsArray($responseArray, 'Response is not valid JSON');
    
    /* Verify that the response does not have errors */
    $this->assertArrayNotHasKey('errors', $responseArray, 'Response contains errors');
    $responseContent = $responseArray['data'];

    /* Verify the properties of the response payload per the schema */ 
    $this->assertArrayHasKey('teasers', $responseContent);
    if ($responseContent['teasers']) {
      $this->assertIsArray($responseContent['teasers']);
      for($i = 0; $i < count($responseContent['teasers']); $i++) {
        if ($responseContent['teasers'][$i]) {
          $this->assertEquals('Teaser' , $responseContent['teasers'][$i]['__typename']);
          $this->assertArrayHasKey('title', $responseContent['teasers'][$i]);
          $this->assertNotNull( $responseContent['teasers'][$i]['title']);
          $this->assertIsString( $responseContent['teasers'][$i]['title']);
          $this->assertArrayHasKey('subTitle', $responseContent['teasers'][$i]);
          if ($responseContent['teasers'][$i]['subTitle']) {
            $this->assertIsString( $responseContent['teasers'][$i]['subTitle']);
          }
          $this->assertArrayHasKey('url', $responseContent['teasers'][$i]);
          $this->assertNotNull( $responseContent['teasers'][$i]['url']);
          $this->assertIsString( $responseContent['teasers'][$i]['url']);
        }
      }
    }
  }
}
\end{lstlisting}